# ALMA Newsletter
September 2010

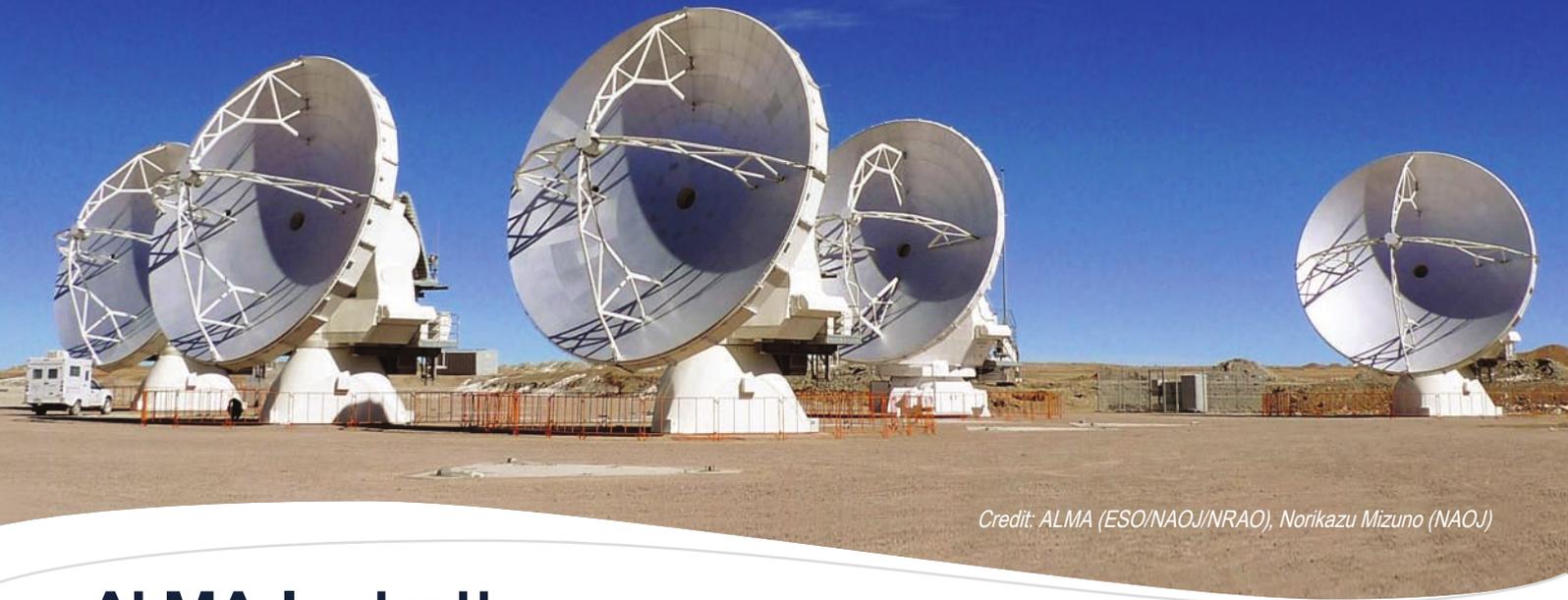

*Credit: ALMA (ESO/NAOJ/NRAO), Norikazu Mizuno (NAOJ)*

## ALMA In-depth

### Sense and sensitivity - how ALMA's receivers work

by **Tom Wilson**, Head of the Radio, Infrared and Optical Sensors Branch at the U.S. Naval Research Laboratory in Washington, D.C., **Rainer Mauersberger**, ALMA scientist and **Antonio Hales**, ALMA Scientist

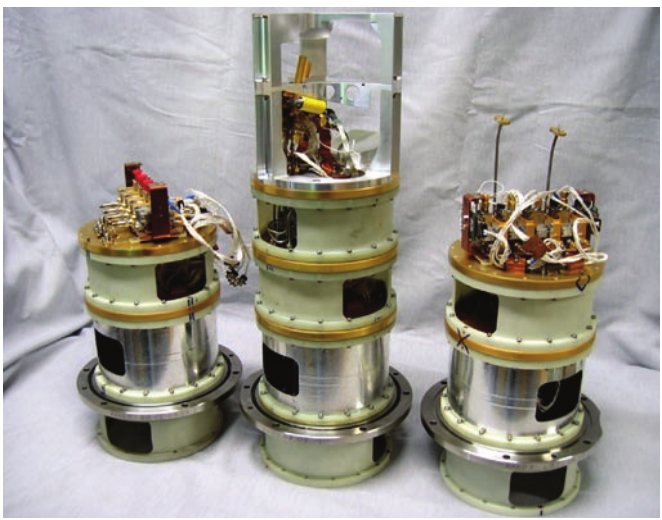

*Fig.1 A photo of three Band 7 prototype front ends in various stages of construction. These are superconducting mixers (SIS) that efficiently convert the power at the sky frequency to a much lower intermediate frequency (IF) where the power is amplified greatly. The advantage of the SIS mixers is that these have contribute an extremely low noise, give rise to only very small losses in the input power and require only a small amount of local oscillator power.*

In previous articles, we described how electromagnetic waves emitted from objects in the sky are collected by the ALMA antennas (Anatomy of ALMA), and how they are combined in order to produce images. Before these images can be processed, they are picked up by the antennas and concentrated by the large main mirror and a smaller secondary mirror in the so called focal point of each antenna. In order to process the data they must be first converted to electromagnetic waves of a lower frequency and amplified. This is the role of the ALMA receivers. In principle they work like a normal FM receiver, but at much higher frequencies. Here we describe how they work and what makes them special.

The receiver systems used for ALMA are the most sensitive built so far. There will be up to 10 different receivers per antenna (also known as ALMA Receiver Bands 1 to 10 ), each of which is sensitive to a specific wavelength (or frequency). By offsetting the telescope and tilting the subreflector one can select which receiver shall be used. Each receiver is equipped with a feed horn, that funnels the energy that is coming from the sky via the antenna into the receiver with minimal losses. It is difficult to amplify and process the high frequency radiation collected by the ALMA antennas. This is why the concept of mixing is employed in a first stage. This is the same principle as in

# ALMA Newsletter
September 2010

*Credit: ALMA (ESO / NAOJ / NRAO), Carlos Padilla*

## ALMA In-depth

a commercial FM receiver: The incoming (sky) signals are combined with an artificial oscillator signal of similar frequency. This produces a beat signal with a frequency that is basically the difference between the sky and the oscillator signal, but still contains all the information received from the celestial source. For obvious reasons, this stage is called the mixer. All ALMA receivers produced so far use superconducting components, the so called SIS (superconductor-insulator-superconductor) diodes, as mixers. The performance of these mixers was presented in ALMA Newsletter #2.

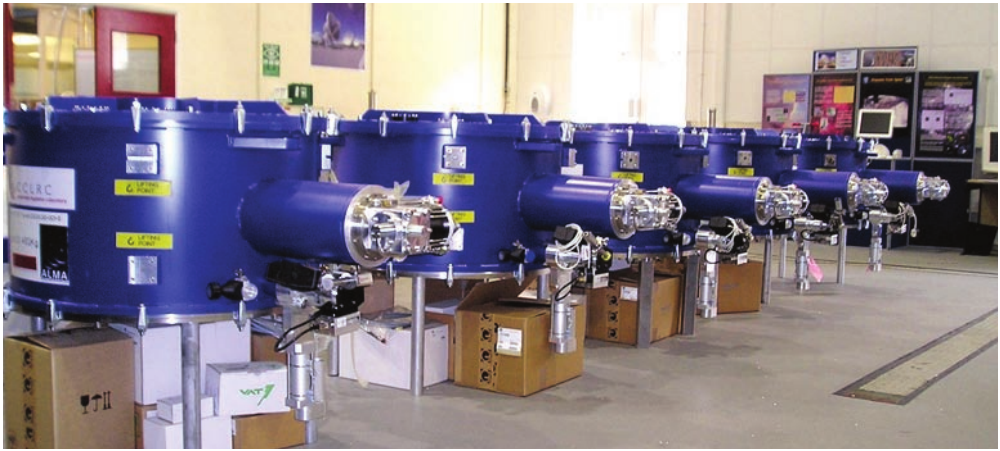

*Fig.2 A photo of the cryostat that houses the SIS mixers and amplifiers. The SIS mixers must be cooled to 4 degrees above absolute zero to function optimally. The amplifiers following the SIS mixers are not cooled to such a low temperature, but these are also extremely low noise systems.*

Briefly, the Front Ends can be described in the following. There are three essential items that contribute to the superior sensitivity of ALMA. These are: (1) the high, dry site, (2) the large collecting area and (3) the receiver system. The ALMA project has devoted a great deal to maximize the sensitivity contribution from the first two items. These include the Chajnantor site and the collecting area provided by 66 antennas that allow operation to 1 Tera Hertz. In the optical and near infrared (near IR) wavelength range, the sensitivity of astronomical detector systems has improved by about a factor of 25 since the 1960's. The sensitivity of millimeter and sub-millimeter astronomical receivers has improved by at least this factor over the same time. In the last few years, receiver sensitivity has improved by another factor of 2 to 4, due to the ALMA development program.

The optical and near IR detectors convert the incident radiation to electrical current in an efficient way. The millimeter and sub-millimeter (mm/sub-mm) receivers must fulfill a number of additional requirements beyond those that apply to the optical and near infrared detectors, that is, the mm/sub-mm receivers must also preserve additional properties of the incident radiation. The most important of these additional properties is the relative arrival time of this radiation at each antenna, that is, the relative phase. In addition, the polarization properties must be accurately recorded. All of this must be done without corrupting the properties of the radiation, or adding more than a small amount of noise power. To accomplish this, the incident radiation is shifted to a longer wavelength (lower frequency) in the first element of the receiver (in a process called down-conversion). Then the power level of the radiation is increased by an enormous factor. Then this output is transferred to the Technical Building at the Array Operations Site (AOS), where the radiation is processed further to produce images.

The noise added by radio astronomy receivers is measured in temperature units, Kelvins. This is universally used as a measure of power for a unit bandwidth. Thus above 200 GHz in the

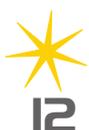

# ALMA Newsletter
September 2010

*Credit: ALMA (ESO / NAOJ / NRAO), Carlos Padilla.*

## ALMA In-depth

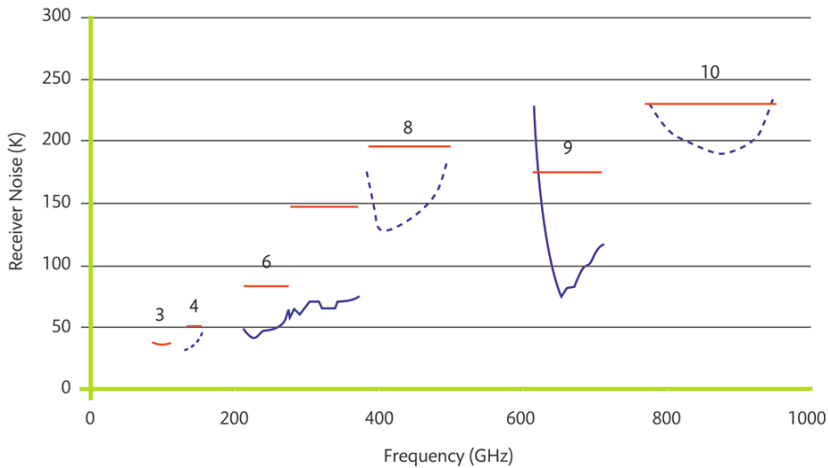

*Fig. 3 Noise temperatures for the ALMA Receiver Bands, on the vertical axis plotted against receiver frequency, on the horizontal axis. The solid horizontal lines mark the specifications for ALMA receiver Bands 3, 4, 6, 7, 9 and 10 (Band 7 is not labeled). The dashed lines under Band 4, 8 and 10 indicate the present status which is in the development phase. The solid irregular lines under Bands 6, 7 and 9 show the present status of laboratory measurements of the receiver noise temperatures for these Bands. In wavelength these ALMA receivers cover a range from 3.5 millimeters to 0.3 millimeters. (adapted from a plot made by J. Webber, NRAO). Not shown on this plot are Receiver Band 5 (at about 180 GHz), which will be installed on six antennas, and Receiver Bands 1(about 30 GHz) and 2 (about 80 GHz), which will be built and installed at a later time.*

plot of Fig. 3, the smooth horizontal line at about 70 K is the ALMA specification, while the uneven line starting at about 50 K is the actually measured receiver performance. It is interesting to compare the receiver noise with noise from the earth's atmosphere. The ideal situation is for a receiver mounted on a lossless antenna outside the earth's atmosphere. We can calculate the noise added to an earth-bound receiver to reach such an ideal. For an ALMA antenna on a very dry site at elevation 5 km, the earth's atmosphere would add 10 K to the noise of a Band 6 receiver. In this situation, the atmosphere adds about 20% to the measured noise of the Band 6 ALMA receiver. Thus, the sensitivity of millimeter receivers, such as Band 6, could still be improved somewhat, but for sub-millimeter receivers, such as Band 9, the earth's atmosphere adds a substantial amount of noise to the receiver contribution. Thus, ALMA receivers are close to ideal systems, especially in the sub-millimeter wavelength range where the earth's atmosphere contributes more noise.

It is worthwhile to present some of the principles needed to achieve the receiver noise temperatures of ALMA receivers. First, the receivers are superconducting devices, so-called Superconducting-Insulating-Superconducting (SIS) devices. The SIS is cooled to 4 Kelvin. The principle is that a flow of electrical current is hindered by the insulating layer. For a given receiver setting, the current will flow when the astronomical signal shines on the device. The SIS device has three useful properties: (1) converting radiation from the sky to electrical current, (2) by combining the frequency of the incoming astronomical signal with the one generated at the AOS building, converting the sky frequency to a much lower frequency (down-conversion), and (3) adding a minimum amount of noise to the astronomical signal. A specific example can be given for Band 7. The astronomical frequency could be, as an example, between 275 and 373 GHz (0.87 millimeters). If the astronomical signal of interest

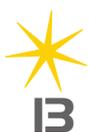

# ALMA Newsletter
September 2010

## ALMA In-depth

is at 345 GHz, this will be converted to 8 GHz (about 4 centimeters wavelength) with the signal properties such as amplitude, phase and sense of polarization preserved. The signal is then amplified. After amplification, the signal (plus a small amount of noise) is sent to the ALMA back end for transmission to the AOS building where it is combined with the outputs of the other ALMA antennas. Once combined, the outputs from 66 antennas are processed to produce an image.

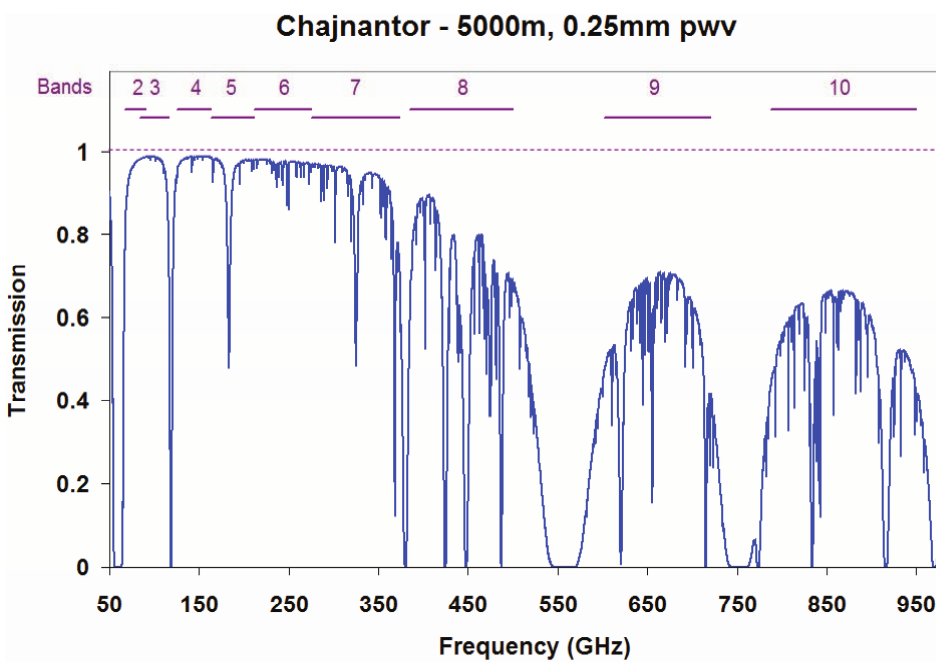

Fig. 4 On the left vertical axis is the transmission of the earth's atmosphere, on the right axis is the noise contributed by the earth's atmosphere. A transmission of 1 means that all the incoming light passes through the atmosphere without any attenuation. Zero percent transmission means that all the incoming radiation is absorbed by the atmosphere. This polt is for the ALMA site, at an elevation of 5 km. The total amount of precipitable water vapor (pwv) above the site can be as low as 0.25 mm This small amount ensures that the effect of the earth's atmosphere on system noise is a minimum.

Due to their nature, the SIS mixers contribute just a very small amount of electronic noise to the signal received, and give rise to only a small loss of the power entering the feed horn. As an example, if we are interested in a frequency range between 340 and 348 GHz in the sky, the astronomer or telescope operator first selects the corresponding band 7 receiver by offsetting the telescope and tilting the subreflector via the telescope control system. The power entering at 340 to 348 GHz is converted to a frequency of 4 to 12 GHz by the SIS mixer. The portion of sky frequencies to be measured is selected by the Local Oscillator (LO). This is an extremely stable, narrow band signal which is sent to all antennas. The power is subsequently increased by a vast amount, to 1 milli Watt, which is a power that can easily be processed.

Following the SIS mixer is another set of mixers that shift the single 4 to 12 GHz band into four bands, each from 2 to 4 GHz. This is done to allow identical electronics for the following stages. In each of those steps the information content of the signals (i.e. the relative frequencies, the relative intensities and, most important the phase, i.e. the timing of each wave) must be preserved. These 2 GHz wide bands, which are still analogue signals, are then digitized in the antennas. The digitized outputs are formatted to insure that no information is lost. These signals are now coded on top of signal of optical light, and transmitted on optic fibers, which are buried in the ground, to the Technical Building (TB) at the ALMA Operations Site (AOS) for further processing, needed to produce

# ALMA In-depth

images. This process was alluded to in the description of "How ALMA will make images" in the last newsletter.

The preceding paragraph gives a correct, but not very informative description of how ALMA will function. A perhaps more enlightening version makes use of analogies. This is given in the following.

In Fig. 6 is a schematic of a single radio receiver of the type used in ALMA. These are referred to as heterodyne receivers. The system shown on the left has a box labeled "RF Amp". This is not present in the ALMA receiver Bands used so far but would be used in the lowest frequency receiver, Band 1, which will be working at such a low frequency that direct amplification is still possible. In ALMA, the final stage is not demodulation (detection with a square law device, that is, the output is the square of the input) but rather further processing of signals in the correlator. On the right are a series of sketches that show the time and frequency behavior of the signal voltage at each stage of the reception process. The topmost line is the signal received from the sky frequency (in the above example from 275-373 GHz). The time behavior shows very rapid variations. The next line shows the behavior in time and frequency if one would amplification at the sky frequency in a rather narrow range of frequencies. Here the time variations of the sky frequency input, are slower since the time variations are inversely proportional to the frequency range filtered out. On the next line, the center frequency of the input

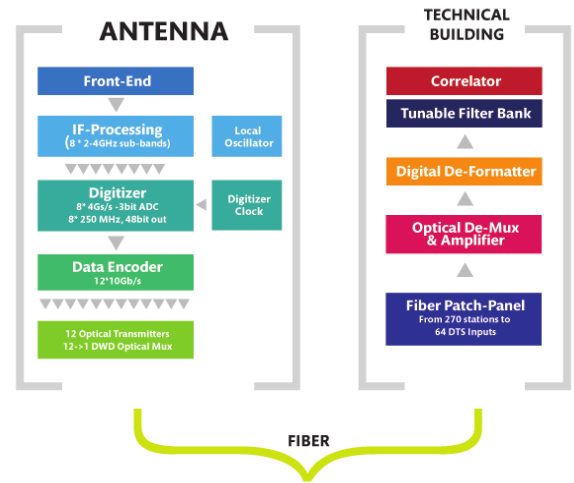

*Fig. 5 A schematic of the ALMA back end design. ALMA electronics in the antenna (on the left) to the Technical Building (TB) at the ALMA Operations Site (AOS). The front end converts the input from the antenna to an IF frequency with four sub-bands of 2 GHz bandwidth for each of two polarizations. The output of each sub-band is digitized at a rate of 4 giga samples per second. This is then encoded, shifted to the optical and then transmitted to the AOS TB where the optical transmission is input to the 32 tunable filter banks (FB). The FB's are used to select specific bands of up to 2 GHz width within the 8 GHz band selected by the front end.*

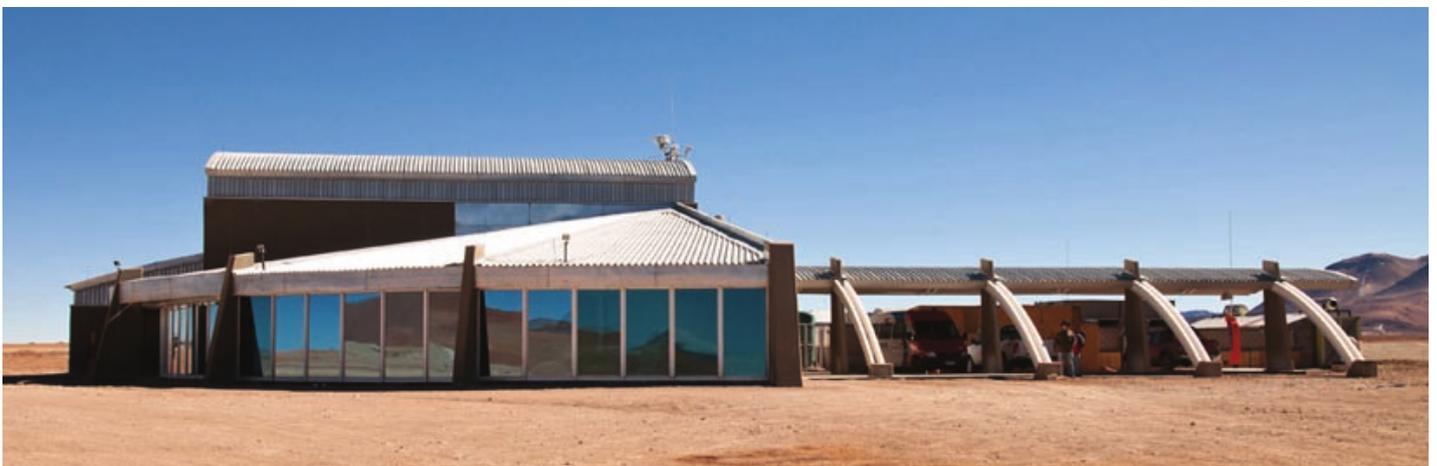

*A photo of the Technical Building at the AOS. The outputs from all ALMA antennas are brought to this facility for further processing.*

# ALMA In-depth

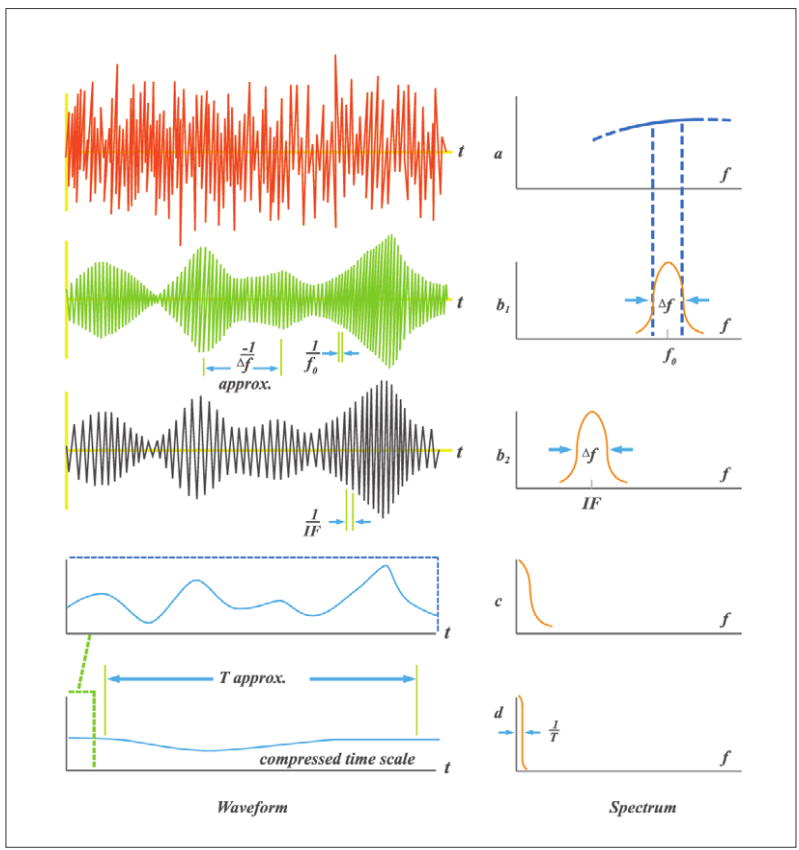

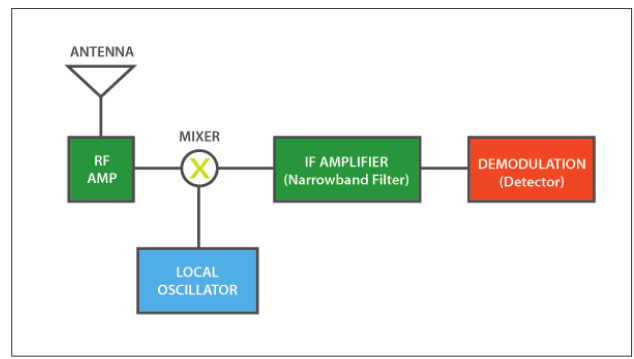

is lowered by the action of the mixer (in our example to a frequency range between 4 and 12 GHz). This is shown in the even slower oscillation time. The lowest two lines show the effect of the detection. There is an analogy for ALMA in that the variation of fringes in the correlation of two antennas is much slower than the variation of sky or IF frequencies.

One can consider a two element interferometer to understand the more complex ALMA system. On the left is a sketch of two elements. In the receivers, located after the antennas, there are a series of narrowly spaced parallel lines. These are frequency bands of 2GHz or less. This allows interferometric spectral line observing and also for continuum observation, since narrower frequency bands make signal processing simpler.

In the method shown, each of the antennas must be equipped with an identical set of filters. For a large system such as ALMA this would be a very formidable task. There is an alternative however (this is the system actually used for ALMA). In this approach, the outputs from the two antennas are sampled at equal times. These samples are shifted relative to each other in time and then multiplied. This process is referred to as correlation. When the correlated products are converted from time to frequency the results are completely equivalent to the setup with two sets of filters. However, the extension using this method to many antennas is rather simple. For the user of ALMA the complex scheme of amplifying, downconverting and processing of astronomical data will be very transparent (if he or she wishes) since the Observing Tool and the data reduction pipeline are designed to translate the scientific goals of ALMA users into system configurations and to convert the visibilities measured into image cubes.

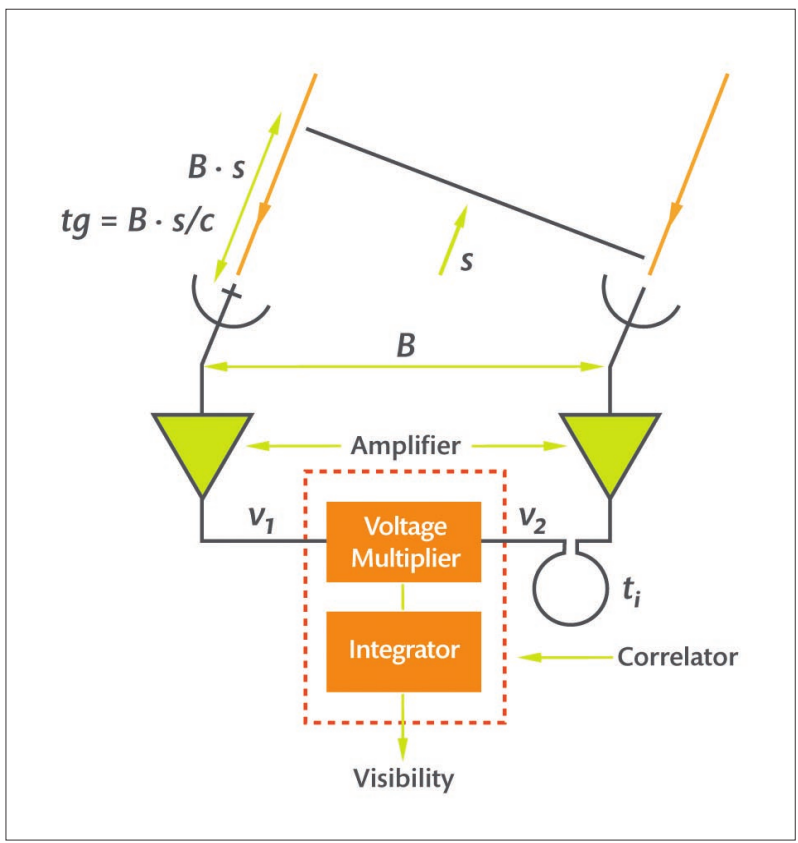